# Global organization of metabolic fluxes in the bacterium, *Escherichia coli*

E. Almaas[1], B. Kovács[1,2], T. Vicsek[2], Z. N. Oltvai[3], A.-L. Barabási[1]

[1]Department of Physics, University of Notre Dame, Notre Dame, IN 46556, USA

[2]Biological Physics Department and Research Group of HAS, Eötvös University, H-1117 Budapest, Hungary

[3]Department of Pathology, Northwestern University, Chicago, IL 60611, USA

**Cellular metabolism, the integrated interconversion of thousands of metabolic substrates through enzyme-catalyzed biochemical reactions, is the most investigated complex intercellular web of molecular interactions. While the topological organization of individual reactions into metabolic networks is increasingly well understood[1-4], the principles governing their global functional utilization under different growth conditions pose many open questions[5-7]. We implement a flux balance analysis[8-12] of the *E. coli* MG1655 metabolism, finding that the network utilization is highly uneven: while most metabolic reactions have small fluxes, the metabolism's activity is dominated by several reactions with very high fluxes. *E. coli* responds to changes in growth conditions by reorganizing the rates of selected fluxes predominantly within this high flux backbone. The identified behavior likely represents a universal feature of metabolic activity in all cells, with potential implications to metabolic engineering.**

To identify the interplay between the underlying topology[1-3] of the *E. coli* K12 MG1655 metabolic network and its functional organization, we focus on the global features of potentially achievable flux states in this model organism with a fully sequenced and annotated genome[13,14]. In accordance with flux-balance-analysis (FBA)[8-12], we first identified the solution space (i.e., all possible flux states under a given condition) using constraints imposed by the conservation of mass and the stoichiometry of the reaction system for the reconstructed *E. coli* metabolic network[8-12]. Assuming that cellular metabolism is in a steady state and optimized for the maximal growth rate, FBA allows us to calculate the flux for each reaction using linear optimization[8-11], providing a measure of each reaction's relative activity. As previous experiments of Pallson and collaborators have shown[8-10], the steady state and optimality approximations offer experimentally verifiable predictions on the flux states of the cell. However, under any condition there are expected differences as well, some coming from the fact that there are transient effects and that the flux distribution is not always exactly optimal[12]. A striking feature of the obtained flux distribution is its overall inhomogeneity: reactions with fluxes spanning several orders of magnitude coexist under the same conditions. For example, under optimal growth conditions in a glutamate rich culture the dimensionless flux of the succinyl-CoA synthetase reaction is 0.185 while the flux of the aspartate oxidase reaction is four orders of magnitude smaller, with a value of 2.2 x $10^{-5}$ in dimensionless units (the flux vector is normalized to unity). To characterize the coexistence of such widely different flux values, in Fig. 1a we plot the flux distribution for active (non-zero flux) reactions of *E. coli* grown in a glutamate- or succinate rich substrate. The distribution is best fitted with a power law with a small flux constant, indicating that the probability that a reaction has flux $\nu$ follows $P(\nu) \sim (\nu + \nu_0)^{-\alpha}$, where the constant is $\nu_0 = 0.0003$ and the flux exponent has the value $\alpha = 1.5$. The observed power-law is consistent with published experimental data as well. Indeed, the flux distribution obtained from the measured fluxes of the central metabolism of *E. coli* [15] is best fitted with the power-law form (Fig. 1d). As the central metabolism is characterized by high fluxes, the small flux saturation seen in Fig. 1a is absent from this data.

To examine whether the observed flux distribution is independent of the environmental conditions, we mimicked the influence of various growth conditions by randomly choosing 10%, 50% or 80% of the 96 potential substrates that *E. coli* can consume (the input substrates are listed in Table S2, Supplementary Material). Optimizing the growth rate, we find that the power law distribution of metabolic fluxes is independent of the external conditions (Fig. 1b). As the metabolic activity of *E. coli* frequently deviates from the optimal growth state under variable growth conditions[10,12], we inspected if the wide flux distribution is present in non-optimal states as well. For this we implemented a "hit-and-run" method[16,17] that randomly samples the full solution space, allowing us to calculate the flux for each reaction in 50,000 distinct non-optimal states. While the obtained average flux distribution is consistent in shape and flux ranges with those obtained by the optimal FBA, the flux exponent is somewhat larger ($\alpha = 2$, Fig. 1c), and the quality of the scaling is slightly weaker. Interestingly, many individual non-optimal states (Fig 1c, inset) are consistent with an exponent $\alpha = 1$, in accord with the experimental results (Fig 1d), supporting the prediction[10,12] that these organisms may not have achieved optimality. These findings imply that the observed flux distribution is a generic feature of flux conservation[18] on a scale-free network[19], being independent of the optimal or non-optimal nature of the growth rate or the growth conditions. The exponent, however, may depend on the organism's position in the solution space.

An exponentially decaying distribution (e.g. a Gaussian) would predict that under a given condition most reactions are characterized by comparable fluxes. In contrast, the identified power law flux distribution suggests a highly uneven utilization: the vast majority of reactions have quite small fluxes, coexisting with a few reactions with very high flux values. Hence, the biochemical activity of the metabolism is dominated by several "hot" reactions, embedded into a network of mostly small flux reactions. The fact that the observed flux exponent $\alpha$ is less or equal to 2 implies that the first and higher moments of the flux distribution are not defined from a mathematical perspective (they diverge for an infinite system). The divergence of the average flux $\langle v \rangle$ (first moment) indicates that no flux value can be designated as characteristic for the system. While $\langle v \rangle$ can be calculated numerically by averaging over the flux of all reactions, providing $\langle v_g \rangle = 0.0171$ and $\langle v_s \rangle = 0.0173$ for glutamate and succinate uptakes respectively, these values are by no means indicative of the metabolic reactions' overall activity. Indeed, the vast majority (86% for glutamate and 89% for succinate uptake) of reactions have a smaller flux than this average, while a few reactions display orders of magnitude higher activity.

The observed flux distribution is compatible with two quite different potential local flux structures. A homogeneous local organization would imply that all reactions producing (consuming) a given metabolite have comparable fluxes. A more delocalized "hot backbone" is expected, however, if the local flux organization is heterogeneous, such that each metabolite has a dominant source (consuming) reaction (Fig. S7, Supplementary Material). To distinguish between these two scenarios for each metabolite $i$ produced (consumed) by $k$ reactions, we define the measure[20]

$$Y(k,i) = \sum_{j=1}^{k} \left( \frac{\hat{v}_{ij}}{\sum_{l=1}^{k} \hat{v}_{il}} \right)^2 , \qquad (1)$$

where $\hat{v}_{ij}$ is the mass carried by reaction $j$ which produces (consumes) metabolite $i$. If all reactions producing (consuming) metabolite $i$ have comparable $\hat{v}_{ij}$ values, $Y(k,i)$ scales as $1/k$. If, however, a single reaction's activity dominates Eq. (1), we expect $Y(k,i) \sim 1$, i.e., $Y(k,i)$ is independent of $k$. For the *E. coli* metabolism optimized for succinate and glutamate uptake we find that both the *in* and *out* degrees follow $Y(k,i) \sim k^{-0.27}$ (Fig. 2a), representing an intermediate behavior between the two extreme cases. This indicates that the large-scale inhomogeneity observed in the overall flux distribution is increasingly valid at the level of the individual metabolites as well: the more reactions consume (produce) a given metabolite, the more likely it is that a single reaction carries the majority of the flux. Such inhomogeneity is obvious e.g., for Flavin Adenine Dinucleotide (FAD), whose production (consumption) is dominated by only one of the 2 (3) contributing reactions (inset of Fig. 2a), the high flux reactions being catalyzed by Succinate dehydrogenase complex, EC 1.3.5.1 (Succinate dehydrogenase, EC 1.3.99.1). We find that $Y(k,i)$ scales in a similar fashion when *E. coli* is grown in rich Luria-Bertani medium, as well as in the non-optimized configurations, indicating that the local inhomogeneity is not a unique feature of the optimized state or specific growth condition but represents a generic property of the local flux distribution (see the Supplementary Material for further details).

The local flux inhomogeneity indicates that for most metabolites we can identify a single reaction dominating its production (consumption). This observation can be turned into a simple algorithm, which systematically removes, for each metabolite, all reactions but the one providing the largest incoming (outgoing) flux contribution. The algorithm uncovers the *high flux backbone* (HFB) of the metabolism, a distinct structure of linked reactions that forms a giant component[21] with a star-like topology (see both Fig. 3 and Fig. S12 of the Supplementary Material) including most metabolites produced under the given growth condition. Only a few pathways appear disconnected, indicating that while these pathways are part of the HFB, their end product serves only as the second most important source for some other HFB metabolite. Of note, groups of individual HFB reactions largely overlap with the traditional, biochemistry-based partitioning of cellular metabolism: e. g., all metabolites of the citric-acid cycle of *E. coli* are recovered, and so are a considerable fraction of other important pathways, such as those being involved in histidine-, murein- and purine biosynthesis, to mention a few. The HFB captures the subset of reactions which dominate the activity of the metabolism. As such, it offers a complementary approach to elementary flux mode analyses[22-24], which successfully capture the available modes of operation for smaller metabolic sub-networks.

As the flux of the individual metabolic reactions depends on the growth conditions, we need to inspect how sensitive the HFB is to changes in the environment. Surprisingly, Fig. 2b and 2c, which record the relationship between the individual fluxes under succinate and glutamate uptake, indicates that only the reactions in the high flux territory undergo noticeable flux changes, while the reactions within the intermediate and low flux region remain virtually unaltered (a small shift, however, can be observed due to the fact that there is a 41% increase in biomass production in glutamate- compared to succinate rich media). The observed flux changes correspond to two types of events. First, certain pathways are turned off completely (type I reactions) having zero flux under one growth condition and high flux in the other. These reactions are shown as symbols along the horizontal and vertical axis in Fig. 2b. In contrast, other reactions remain active but display orders of magnitude shifts in fluxes under the two different growth conditions (type II reactions). Of note, with two exceptions, these drastic type II changes are limited to the HFB reactions. The same phenomenon is predicted when we inspect the transition from glucose to succinate uptake, or for transitions between various random uptake conditions (see the Supplementary Material for additional details).

To test the generality of this finding, we mimicked the effect of various growth conditions by randomly choosing 50% of the potential input substrates, measuring in each input configuration the flux for each reaction. For each reaction the average flux ($v$), as well as the standard deviation ($\sigma$) around this average, was determined by averaging over 5000 random input conditions. As Fig. 2c depicts, for small fluxes all reactions closely follow a straight line (corresponding to $\sigma \sim \nu$), supporting the earlier finding that the small fluxes remain essentially unaltered as the external conditions change. For the high flux reactions, however, there are noticeable deviations from this line, indicating significant flux variations from one external condition to the other. A closer inspection of the flux distribution shows that the reactions on the $\sigma \sim \nu$ curve all have a clear unimodal flux distribution (Fig. 4a and 4c), indicating that shifts in growth conditions lead to only small changes (within $\sigma$) of their flux values. In contrast, the reactions deviating from the $\sigma \sim \nu$ curve display a bi- or trimodal distribution, indicating that under different growth conditions they exhibit several discrete and quite distinct flux values (Figs. 4b and 4d). Therefore, Figs. 2c, 3 and 4 offer valuable insights on how *E. coli* responds to changes in growth conditions: it (de)activates certain metabolic reactions among the HFB metabolites in novel ways without altering the identity of the major pathways that participate in the backbone. For example, switching from a glutamate to a succinate substrate turns off the vitamin B6 biosynthesis (Type I change), and reconnects in new ways several metabolic substrates, like those of the Coenzyme A and NAD biosynthesis pathways as well as the respiration pathways, and modifies the use of the TCA cycle (Fig. S12, Supplementary Material). Apart from minor changes, the utilization of the other pathways remains unaltered. These reorganizations result in major discrete changes in the fluxes of the HFB reactions.

The power-law distribution of metabolic fluxes within the *E. coli* metabolism indicates a highly uneven utilization of the underlying metabolic network topology. While wide flux differences among various pathways are known from individual experimental observations[15,25-27], we find that they are part of a scale-invariant continuum, following a scaling law. The uneven flux utilization is present at both the global level (Fig. 1), and at the level of the individual metabolites (Fig. 2a). This observation allows us to automatically uncover the high flux backbone of the metabolism and could provide significant insights into metabolic organization and regulation as well as offering valuable inputs for metabolic engineering.

The observation and theoretical prediction of a power-law load distribution in simple models (Ref. 18 and Supplementary Material), as well as the presence of a power law in both the optimal and non-optimal flux states, suggests that the metabolic flux organization is a direct consequence of the network's scale-free topology. As all organisms examined to date are characterized by a scale-free[19] metabolic network topology[1], the observed scaling in the flux distribution is likely not limited to *E. coli*, but characterizes all organisms from eukaryotes to archaea. As FBA is available for an increasing number of prokaryotic- and eukaryotic organisms, this prediction could be verified both experimentally and theoretically in the near future. Hence, the observed uneven local and global flux distribution appears to be rooted in the subtle, yet generic, interplay of the network's directed topology and flux balance, channeling the numerous small fluxes into high flux pathways. The dependence of the scaling exponents characterizing the flux distributions on the nature of the optimization process, as well as the experimentally observed exponent, may serve as a benchmark for future structural and evolutionary models aiming to explain the origin, the organization and the modular structure[3,4,28,29] of cellular metabolism.

## Methods

**Flux balance analysis (FBA)** Starting from a stoichiometric matrix of the MG1655 [8,9] strain of *E. coli*, containing 537 metabolites and 739 reactions, the steady state concentrations of all the metabolites satisfy

$$\frac{d}{dt}[A_i] = \sum_j S_{ij} v_j = 0, \qquad (2)$$

where $S_{ij}$ is the stoichiometric coefficient of metabolite $A_i$ in reaction $j$ and $v_j$ is the flux of reaction $j$. We use the convention that if metabolite $A_i$ is a substrate (product) in reaction $j$, $S_{ij} < 0$ ($S_{ij} > 0$), and we constrain all fluxes to be positive by dividing each reversible reaction into two "forward" reactions with positive fluxes. Any vector of positive fluxes $\{v_j\}$ which satisfies (2) corresponds to a state of the metabolic network, and hence, a potential state of operation of the cell. We restrict our study to the subspace of solutions for which all components of $\nu$ satisfy the constraint $v_j > 0$ [8]. We denote the mass carried by reaction $j$ producing (consuming) metabolite $i$ by $\hat{v}_{ij} = |S_{ij}| v_j$.

**Random uptake conditions** We choose randomly X%, (where X=10, 50 or 80) of the 89 potential input substrates *E. coli* consumes in addition to the minimal uptake basis. For each of the transport reactions we set the uptake rate to 20 mmol/g DW/h. As there is a very large number of possible combinations of the selected input substrates, we repeat this process 5000 times and average over each realization.

**The "hit-and-run" method**[16,17] We select a set of basis vectors spanning the solution space using singular-value decomposition. Since the reaction fluxes must be positive, the "bouncer" is constrained to the part of the solution space intersecting the positive orthant. We constrain the bouncer within a hypersphere of radius $R_{max}$ and outside of a hypersphere of radius $R_{min} < R_{max}$, where we find that the sampling results are independent of the choices of $R_{min}$ and $R_{max}$. Starting from a random initial point inside the positive flux cone in a randomly chosen direction the bouncer travels deterministically a distance *d* between sample points. Each sample point, corresponding to a solution vector where the components are the individual fluxes, is normalized by projection onto the unit sphere. After every $b^{th}$ bounce off the internal walls of the flux cone the direction of the bouncer is randomized.

**High flux backbone:** For each metabolite we only keep the reactions with the largest flux producing and consuming the metabolite. Metabolites not produced (consumed) are discounted. Subsequently, a directed link is introduced between two metabolites A and B if (i) A is a substrate of the most active reaction producing B, and (ii) B is a product of the maximal reaction consuming A. We consider only metabolites which are connected to at least one other metabolite after steps (i) and (ii). For clarity we removed $P_i$, $PP_i$ and ADP. Further details and figures are provided in the Supplementary Material.

**Supplementary information** accompanies the paper on *Nature's* website (http://www.nature.com).

Correspondence and requests for materials should be addressed to A.-L. Barabási (alb@nd.edu).

**Acknowledgements**
We thank E. Ravasz, A. Vazquez, and S. Wuchty for useful discussions and B. Pallson and S. Schuster for outstanding comments on the manuscript. Research at Eötvös University was supported by the Hungarian National Research Grant Foundation (OTKA), while work at the University of Notre Dame and at Northwestern University was supported by the U.S. Department of Energy, the National Institute of Health and the National Science Foundation.

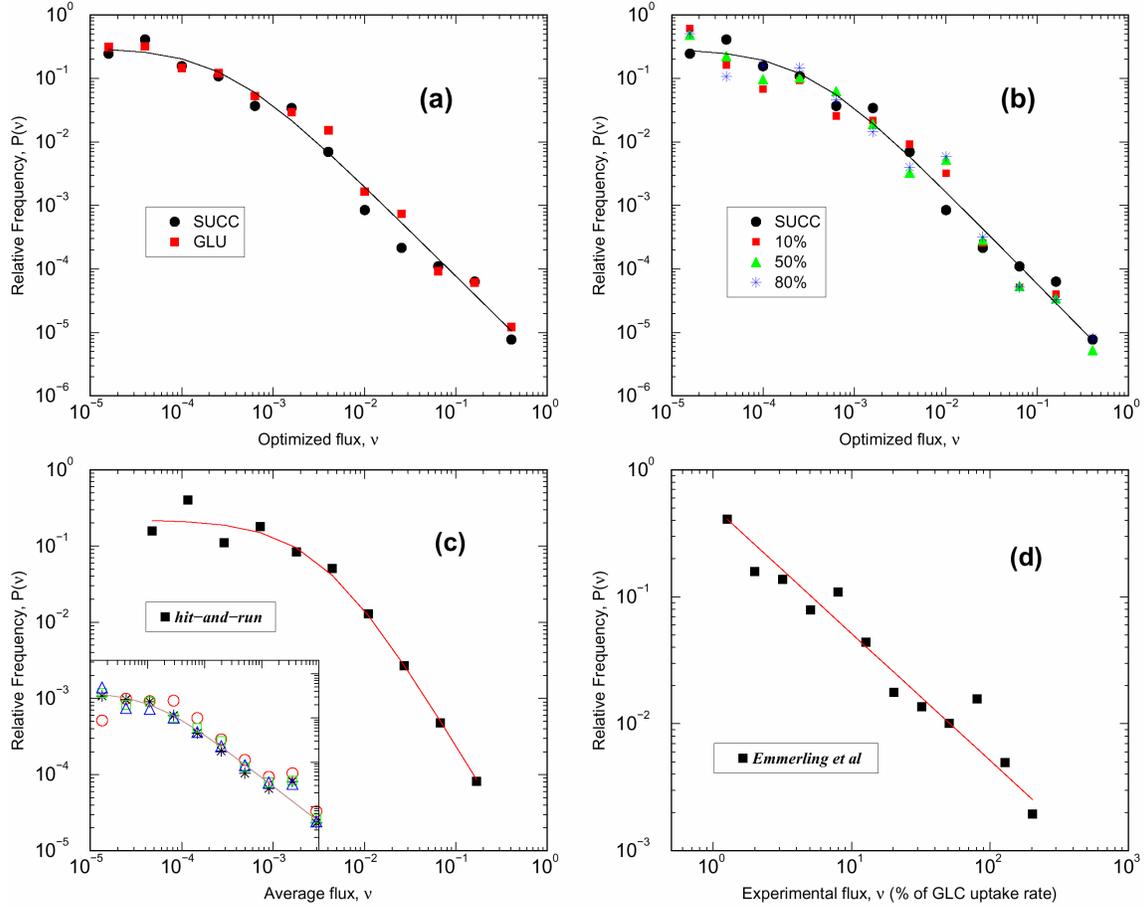

**Figure 1.** Characterizing the overall flux organization of the *E. coli* metabolic network. **(a)** Flux distribution for optimized biomass production on succinate (black) and glutamate (red) substrates. The solid line corresponds to the power law fit $P(\nu) \sim (\nu + \nu_0)^{-\alpha}$ with $\nu_0 = 0.0003$ and $\alpha = 1.5$. **(b)** Flux distribution for optimized biomass on succinate (black) substrate with an additional 10% (red), 50% (green) and 80% (blue) randomly chosen subsets of the 96 input channels turned on. We averaged the flux distribution over 5000 independent random choices of uptake metabolites. The resulting flux distribution can be fitted (solid line) with a power law with parameters $\nu_0 = 0.0004$ and $\alpha = 1.5$. **(c)** Flux distribution from the non-optimized "hit-and-run" sampling method of the *E. coli* solution space. The solid line is the best fit with $\nu_0 = 0.003$ and $\alpha = 2$. The inset shows the flux distribution in four randomly chosen sample points. **(d)** The distribution of experimentally determined fluxes (see Ref. 15) from the central metabolism of *E. coli* displays power-law behavior, with a best fit to $P(\nu) \sim \nu^{-\alpha}$ with $\alpha = 1$.

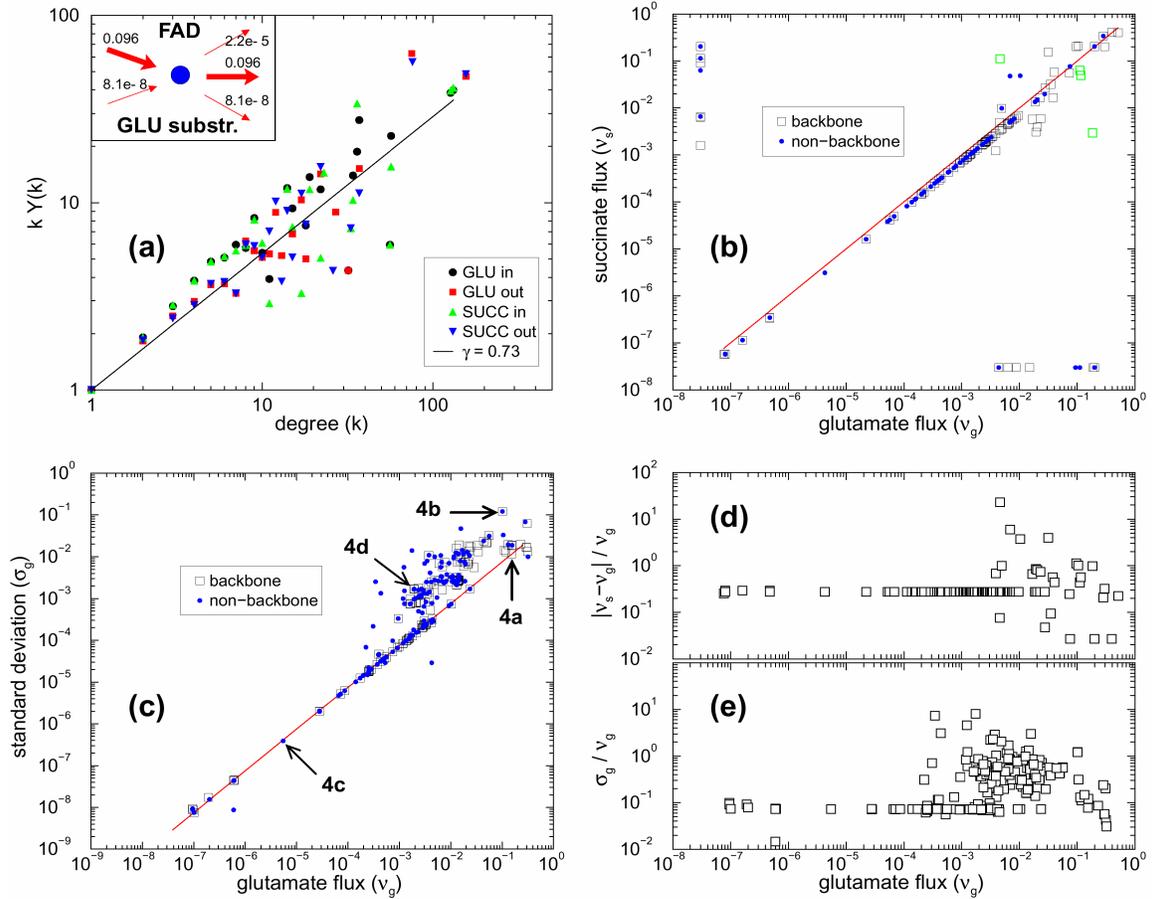

**Figure 2.** Characterizing the local inhomogeneity of the metabolic flux distribution. (**a**) The measured $kY(k)$ shown as function of $k$ for incoming and outgoing reactions, averaged over all metabolites, indicating $Y(k) \sim k^{-0.27}$, as the straight line in the figure has slope $\gamma = 0.73$. **Inset:** The non-zero mass flows $\hat{v}_{ij}$ producing (consuming) flavin adenine dinucleotide (FAD) on a glutamate rich substrate. (**b**) The change in the flux of individual reactions when departing from glutamate (horizontal axis, $v_g$) to succinate (vertical axis, $v_s$) rich conditions. Reactions with negligible flux changes follow the diagonal (solid line). Some reactions are turned off in only one of the conditions (shown close to the coordinate axes). Reactions belonging to the high flux backbone (see Fig. 3) are shown as black squares, while the rest are denoted by blue dots. Reactions for which the direction of the flux is reversed are colored green. (**c**) The absolute value of glutamate flux $v_g$ for 50% randomly chosen input channels (averaged over 5000 realizations) plotted against the standard deviation of the same reaction. The red line, corresponding to $y = 0.075x$, is shown for reference. The numbers 4a – 4d identify the reactions whose distribution is displayed in Fig. 4. (**d**) The relative flux change for the conditions shown in (b), $\Delta v = (|v_s - v_g|/v_g)$, as function of $v_g$. (**e**) The relative fluctuation $\sigma_i / v_i$ per reaction for the conditions shown in (c), again emphasizing that the large changes are limited to high flux reactions.

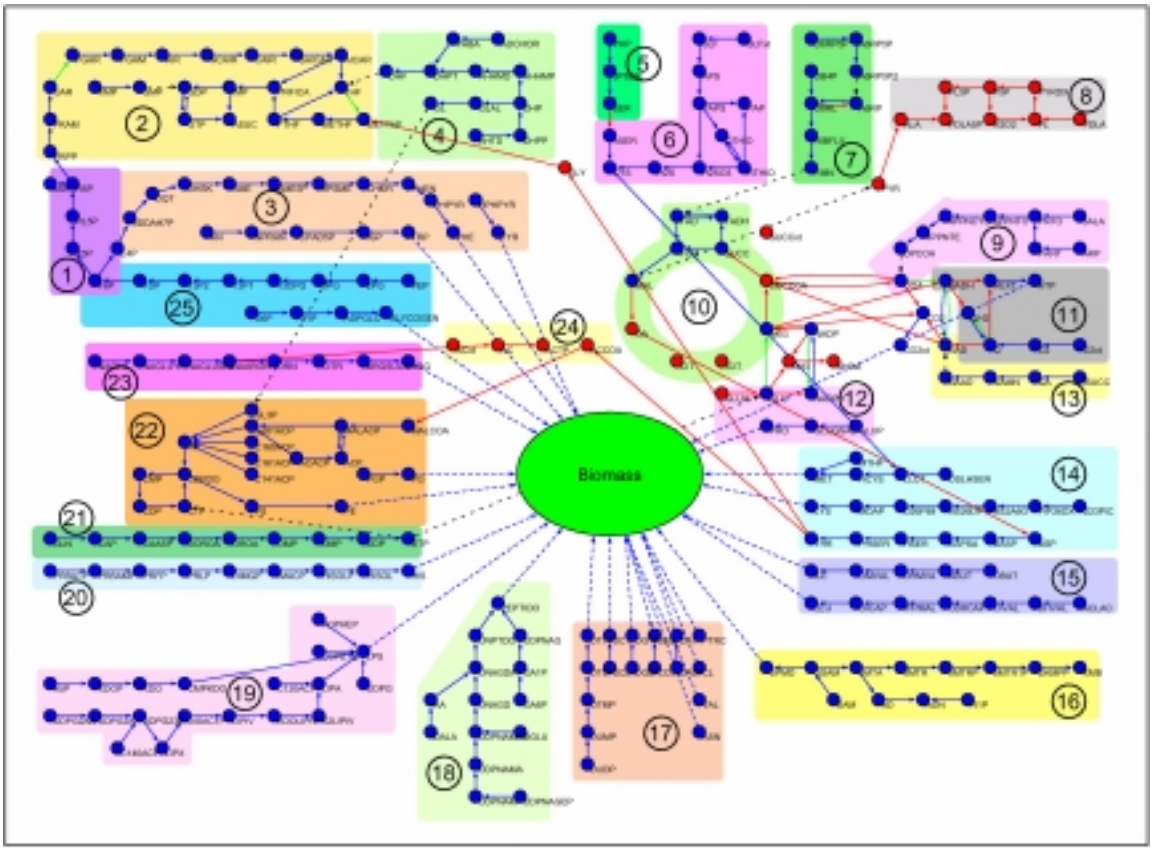

**Figure 3.** High flux backbone (HFB) for FBA optimized metabolic network of *E. coli* on a glutamate rich substrate (see Fig. S12b for succinate rich substrate). We connect two metabolites A and B with a directed link pointing from A to B only if the reaction with maximal flux consuming A is the reaction with maximal flux producing B. We show all metabolites which have at least one neighbor after the completion of this procedure. The background colors demarcate different known biochemical pathways. Metabolites (vertices) colored blue have at least one neighbor in common in glutamate and succinate rich substrates, while those colored red have none. Reactions (edges) are colored blue if they are identical in glutamate and succinate rich substrates, green if a different reaction connects the same neighbor pair and red if this is a new neighbor pair. Black dotted edges indicate where the disconnected pathways (e.g. 4, Folate Biosynthesis) would connect to the cluster via a link that is not part of the HFB. Thus, the red nodes and links highlight the predicted changes in the HFB when shifting *E. coli* from glutamate-rich- to succinate-rich media. Dashed edges indicate links to the biomass growth reaction. The numbers identifying the various biochemical pathways are listed in the Supplementary Material.

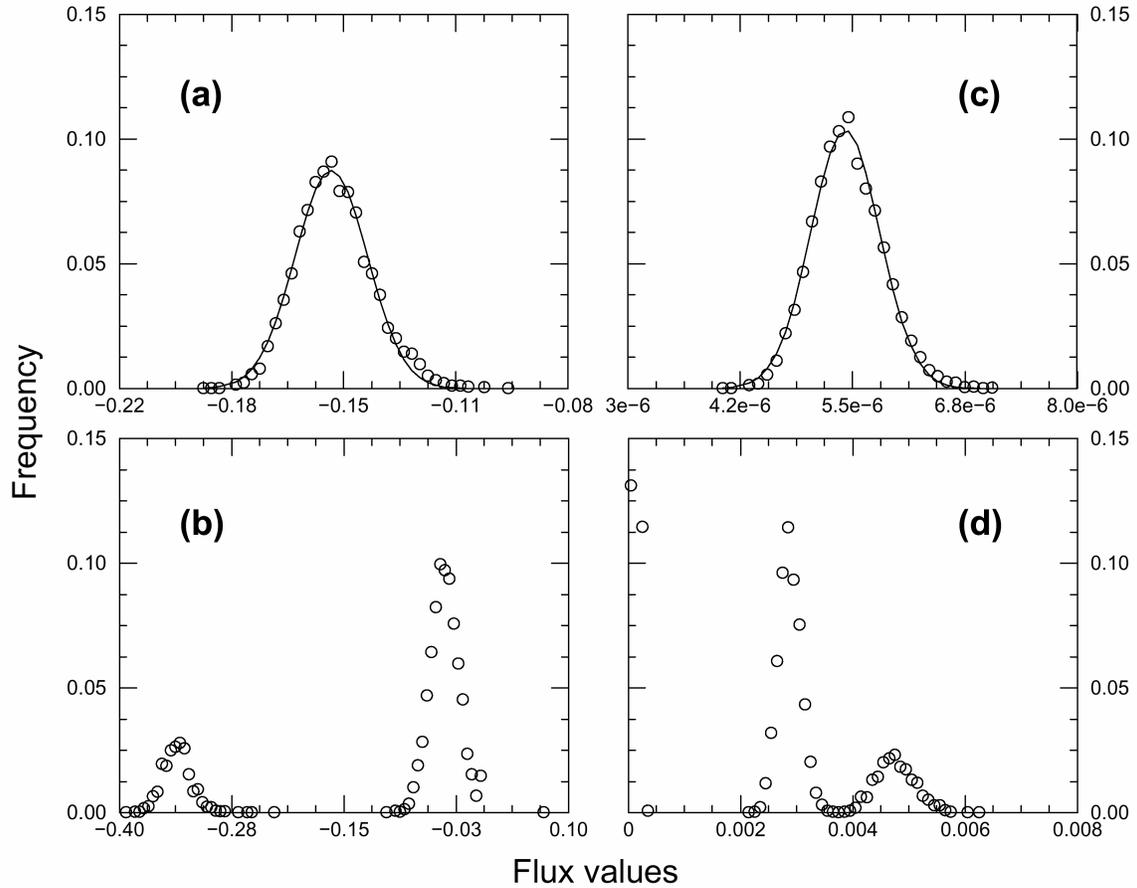

**Figure 4.** The effect of growth conditions on individual fluxes. We show the flux distribution for four select *E. coli* reactions in a 50% random environment (see Fig. 2c). We find that reactions on the $\sigma \sim v$ curve have Gaussian distributions (**a** and **c**) while reactions off this curve have multimodal distributions (**b** and **c**), displaying several discrete flux values under diverse conditions. The solid curves correspond to Gaussians using the calculated ($v, \sigma$) for (**a**) (-0.15, 0.012) and (**c**) (5.4e-6, 3.9e-7). The shown reactions are (**a**) Triosphosphate Isomerase, (**b**) Carbon dioxide transport, (**c**) NAD kinase and (**d**) Guanosine kinase.

15